\documentclass{elsart}
\usepackage{epsfig}

\newcommand{\chib}{\overline{\chi}}
\newcommand{\cbcex}{\langle\overline{\chi}\chi\rangle}
\begin{document}
%
\begin{frontmatter}
  \title{\hfill {\small Aachen, PITHA 97/22}\\ 
\vspace{1cm}
Strongly coupled compact lattice QED \\
    with staggered fermions\\
} 
\author{J.~Cox, W.~Franzki,
    J.~Jers{\'a}k} 
\address{Institut f{\"u}r Theoretische Physik E,
    RWTH Aachen, Germany} 
\author{C.~B.~Lang} 
\address{Institut
    f{\"u}r Theoretische Physik, Karl-Franzens-Universit\"at Graz,
    Austria} 
\author{T.~Neuhaus} 
\address{SCRI, Florida State
    University, Tallahassee, U.S.A.} 
\date{\today}
\begin{abstract}
  We explore the compact U(1) lattice gauge theory with staggered
  fermions and gauge field action $-\sum_P [\beta \cos(\Theta_P) +
  \gamma \cos(2\Theta_P)]$, both for dynamical fermions and in the
  quenched approximation. ($\Theta_P$ denotes the plaquette angle.)
  In simulations with dynamical fermions at various $\gamma \le -0.2$
  on $6^4$ lattices we find the energy gap at the phase transition of
  a size comparable to the pure gauge theory for $\gamma \le 0$ on the
  same lattice, diminishing with decreasing $\gamma$.  This suggests a
  second order transition in the thermodynamic limit of the theory
  with fermions for $\gamma$ below some finite negative value.
  Studying the theory on large lattices at $\gamma = -0.2$ in the
  quenched approximation by means of the equation of state we find
  non-Gaussian values of the critical exponents associated with the
  chiral condensate, $\beta \simeq 0.32$ and $\delta \simeq 1.8$, and
  determine the scaling function.  Furthermore, we evaluate the meson
  spectrum and study the PCAC relation.
\end{abstract}
\end{frontmatter}
\section{Introduction}

Many years of nonperturbative lattice studies have not yet clarified the
properties of the strongly coupled QED.  Concerning the {\em noncompact}
lattice formulation with staggered fermions the properties of the $2^{\rm nd}$
order phase transition between the phase with broken chiral symmetry at strong
coupling and the symmetric weak coupling (Coulomb) phase are still
controversial (see \cite{HaKo94,AzDi96,GoHo97} and references therein). So is
the existence of the conjectured non-Gaussian fixed point which might allow to
avoid triviality of QED in the limit of infinite cutoff. The {\em compact}
strongly coupled QED (U(1) gauge theory) with staggered fermions was found to
have a phase with broken chiral symmetry, too
\cite{AzCr86,SaSe91a,SaSe91b,GaLa91}. But the corresponding phase transition
to the symmetric phase has not been considered of interest for the continuum
theory, as this transition was assumed to be of $1^{\rm st}$ order because of
an observed two-state signal on finite lattices \cite{KoDa87,DaKo88}.

There are two reasons to reconsider the order of the phase transition
in compact QED. Firstly, a suitable modification of the lattice action might
change the order. Such a phenomenon occurs in the pure
compact U(1) lattice gauge theory with the extended Wilson action
\begin{equation}
      S_U = -\sum_P
             \left [\beta \cos(\Theta_P) + \gamma
                 \cos(2\Theta_P)\right ].
\label{S_PURE}
\end{equation}
($\Theta_P$ is the plaquette angle proportional to $F_{\mu\nu}$. Further
details are given in sec.~2.) There the analysis of the decreasing two-state
signal \cite{Bh82,EvJe85} suggested, that the order of the transition line
$\beta = \beta_c(\gamma)$ changes from the $1^{\rm st}$ to the $2^{\rm nd}$
with decreasing $\gamma$.  In the full theory with fermions of bare mass $m_0$
(in lattice units) the two-state signal was found to decrease with decreasing
$\gamma$ for fixed lattice size and $m_0$, too \cite{DaKo88,Ok89}.

Secondly, the recent investigation of the pure U(1) gauge theory with
the extended Wilson action \cite{LaNe94b,JeLa96a,JeLa96b} and the
Villain action \cite{LaPe96} on finite lattices with sphere-like
topology revealed that the two-state signal might in part be due to
finite size effects of topological origin. In some range of couplings,
when the two-state signal is rather weak, it vanishes when the
lattices with sphere-like topology instead of the usual toroidal ones
are used, strongly suggesting $2^{\rm nd}$ order in the thermodynamic
limit. The two-state signal may be related to certain gauge field
configurations, e.g.  monopole loops winding around a toroidal
lattice.  Since the simulations of the full compact QED were performed
on cubic lattices with periodic boundary conditions for gauge fields,
the same phenomenon should be expected also here, quite independently
of the fermions.  Thus it may well be that in the thermodynamic limit
the two-state signal again vanishes and the phase transition is
actually of $2^{\rm nd}$ order.

In this paper we therefore hypothesize that in compact QED with the extended
Wilson action and light staggered fermions a part of the phase transition
surface (parametrized by $\gamma$ and $m_0$) is of $2^{\rm nd}$ order. It
would be challenging to check this conjecture in simulations with dynamical
fermions on lattices with sphere-like topology. Unfortunately, putting
fermions on such lattices is demanding, requiring various preparatory studies,
e.g.  with free fermions. So presently we are not in the position to make such
a direct test.

On the usual toroidal lattices of fixed sizes one can at least investigate,
how the two-state signal depends on $\gamma$ and $m_0$ and compare its size
to that of the pure U(1) gauge theory. Here we report on such a systematic
study of the two-state signal on $6^4$ lattice. The size of the two-state
signal can be conveniently characterized by the difference (``gap'') $\Delta
e_P$ between the mean values $e_P$ of $\cos(\Theta_P)$ in each of the
coexistent states.

A study of the $\gamma$-dependence of $\Delta e_P$ at fixed $m_0$ was made by
Okawa \cite{Ok89} some time ago. He suggested that at $\gamma = -1.3$ and
$m_0=0.1$ the two-state signal is already absent and the transition is
therefore of $2^{\rm nd}$ order there. Our present study with larger
statistics and a modern hybrid Monte Carlo fermion algorithm shows that the
two-state signal persists on the toroidal $6^4$ lattice even at $\gamma =
-1.5$ (simulations at still smaller $\gamma$ are prohibitively expensive).
However, $\Delta e_P$ is smaller than in the pure U(1) theory at $\gamma
\simeq -0.5$, where recent work on spherical lattices \cite{JeLa96a,JeLa96b},
as well as earlier investigations on toroidal lattices \cite{EvJe85} strongly
suggest $2^{\rm nd}$ order.

At fixed $\gamma$ the gap $\Delta e_P$ increases as $m_0$ decreases.  Since in
the continuum limit fermions of finite mass can be obtained only when $m_0
\rightarrow 0$, the value of $\Delta e_P$ at $m_0 = 0$ must be considered. At
$\gamma = -0.2$ we find that $\Delta e_P$, extrapolated to this limit, is of
the same magnitude as in the pure gauge theory at $\gamma = 0$. Numerical
evidence on spherical lattices suggests $2^{\rm nd}$ order in the latter case
\cite{LaNe94b,JeLa96a,JeLa96b}. Thus our results of dynamical fermion
calculations on $6^4$ lattice, described in sec.~3, support the hypothesis
that at some negative $\gamma$ the transition is indeed of $2^{\rm nd}$ order
even for light fermions.

If so, then it is justified to consider the continuum limit and investigate
its physical properties.  We do first steps in this direction, using the
quenched approximation. As described in sec.~4, we have studied on toroidal
lattices at $\gamma = -0.2$ the chiral condensate, susceptibilities, equation
of state and the meson masses. The data have been obtained in Monte Carlo runs
for a simultaneous investigation of the gauge-ball spectrum in the pure U(1)
gauge theory in both phases \cite{CoFr97b}. For that purpose we have
accumulated a large number of gauge field configurations at 17 $\beta$-points on
lattices up to $20^340$ (further details are given in \cite{CoFr97b}). All
these points are close to the phase transition, but outside the tunneling
region where the two-state signal occurs.

\section{Action and phase diagram}

Compact QED with extended Wilson action for gauge field and
staggered fermions on the lattice is defined by the action
\begin{eqnarray}
      S &=& S_\chi + S_U,
       \label{S_WHOLE}\\
      S_\chi &=& \sum_{x,y} \overline{\chi}_x M_{x,y}\chi_{y}\\
      & \equiv & \frac{1}{2} \sum_x \overline{\chi}_x
      \sum_{\mu=1}^4 \eta_{x\mu} (U_{x,\mu} \chi_{x+\mu} -
U^\dagger_{x-\mu,\mu}
      \chi_{x-\mu}) + {m_0} \sum_x \overline{\chi}_x \chi_x ,\\
      S_U &=& -\sum_P
             \left [\beta \cos(\Theta_P) + \gamma
                 \cos(2\Theta_P)\right ].
\end{eqnarray}

Here, $\Theta_P \in [0,2\pi)$ is the plaquette angle, i.e. the
argument of the product of U(1) link variables around a plaquette $P$,
and $\beta$ and $\gamma$ are the single and the double charge
representation couplings, respectively. Taking $\Theta_P =
a^2gF_{\mu\nu}$, where $a$ is the lattice spacing, and $\beta +
4\gamma = 1/g^2$, one obtains for $S_U$ in the limit of weak coupling
$g$ the usual continuum gauge field action $S =\frac{1}{4} \int
d^4xF_{\mu\nu}^2$.  The staggered fermion field $\chi$ corresponds to
four fermion species in the continuum limit.  The lattice model has
global U(1)$\otimes$U(1) chiral symmetry for vanishing fermion bare
mass, $m_0= 0$.  The limit $m_0= \infty$ corresponds to the pure U(1)
gauge theory with extended Wilson action.

Throughout this work we consider the theory on hypercubic lattices
with periodic boundary conditions in all directions except the
antiperiodic boundary condition of the fermion field in one (``time'')
direction. We call such lattices ``toroidal''.

From the accumulated numerical evidence we know that for any fixed
value of $m_0$ this lattice gauge theory has a line of phase
transitions in the $(\beta, \gamma)$-plane between the strong coupling
confinement phase and the weak coupling Coulomb phase.  On finite
toroidal lattices it has a two-state signal (gap $\Delta e_P$) in
$\langle \cos \Theta_P\rangle$ in a broad region of $\gamma$-values,
including $\gamma = 0$ (Wilson action)
\cite{KoDa87,DaKo88,Ok89,AzDi90b,AzCr91,AzBa95}.

\begin{figure}[tbp]
  \begin{center}
    \psfig{file=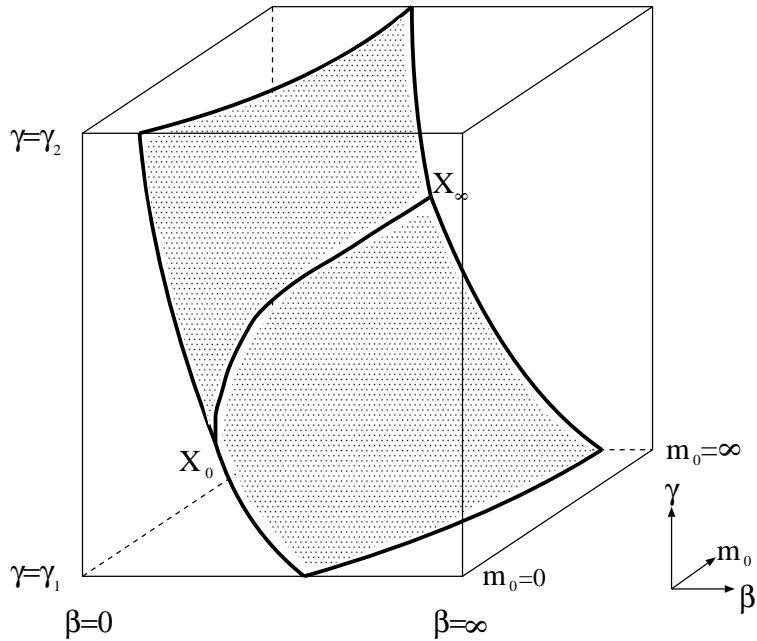,angle=270,width=10cm}
  \end{center}
  \caption{Expected schematic position and nature of the
    confinement-Coulomb phase transition sheet in the three-dimensional phase
    diagram of compact lattice QED with staggered fermions. The sheet of phase
    transitions is of $1^{\rm st}$ ($2^{\rm nd}$) order above (below) the line
    $X_\infty X_0$.  The point $X_0$ might be at $\gamma = -\infty$. The
    $\gamma$ axis extends from large negative $\gamma_1$ to $\gamma_2 \simeq
    0.5$. }
  \label{fig:3dphase}
\end{figure}

These metastability phenomena are similar to those in the pure U(1) lattice
theory.  The recent studies of that theory on spherical lattices strongly
suggest that in spite of non-vanishing $\Delta e_P$ the order changes at
$\gamma = \gamma_0 \simeq 0$, $\gamma_0$ probably being slightly positive, and
is of $2^{\rm nd}$ order for $\gamma \le \gamma_0$ \cite{JeLa96a,JeLa96b}.  In
both theories the two-state signal on toroidal lattices weakens with
decreasing $\gamma$, but is present until the large autocorrelation time at
large negative $\gamma$ makes simulations prohibitively expensive. Thus on
toroidal lattices the two-state signal on the phase transition line cannot be
avoided. However, an analogy with the pure U(1) theory suggests that in the
full theory for any $m_0 > 0$ the transition becomes $2^{\rm nd}$ order at
some {\em finite} negative value of $\gamma$.  Fig.~\ref{fig:3dphase} shows
the conjectured sheet of phase transitions in the thermodynamic limit.

For fixed $\gamma$ (and $\beta$ at the transition) the two-state signal
increases with decreasing $m_0$. Thus the supposed region of $2^{\rm nd}$
order shifts to lower values of $\gamma$.  Whether the order changes at finite
$\gamma$ also at $m_0 = 0$ (as suggested in fig~\ref{fig:3dphase} by the
position of the point ${\rm X}_0$) is a challenging question. 

The continuum limit of the model can be considered at any point of the $2^{\rm
  nd}$ order surface below the line ${\rm X}_\infty {\rm X}_0$, as well as at
the line itself. We expect that for $m_0>0$ this limit will be in the
universality class of the pure gauge theory for the following reason: In the
continuum limit the lattice constant $a$ is determined in physical units (e.g.
by fixing some mass in physical units). The fermion bare mass must be
expressed in such units too,
\begin{equation}
                 m_0 = a\, m^{\rm phys}_0.
\label{BARE_MASS}
\end{equation}
For nonzero fixed $m_0$ the physical fermion mass diverges in the $a
\rightarrow 0$ at nonzero $m_0$, and fermions decouple. Thus for $m_0>0$ we
expect to find the same non-Gaussian universality class as in the pure gauge
theory investigated in \cite{CoFr97b}.

A continuum limit with finite $m^{\rm phys}_0$ requires $m_0\propto a\to 0$.
Furthermore, if a chiral symmetric theory $m^{\rm phys}_0 = 0$ should be
constructed, then $m_0$ must approach zero faster than $a$.

It should be pointed out that, if $\gamma < 0$, the action (\ref{S_WHOLE})
does not satisfy reflection positivity.  Of course, reflection positivity is a
sufficient but not a necessary condition for unitarity, so its absence by
itself does not invalidate the investigation of a lattice model for the
purpose of quantum field theory. However, it makes it more demanding, as
possible sources of unitarity violation must be identified and their scale
determined. If they are on the cut-off level, they do no harm, at least if the
cut-off can be removed.

In our case the fermionic part of the action is reflection positive and the
problem is due to the gauge field action at $\gamma < 0$. Numerical
investigation of that theory on spherical lattices \cite{JeLa96a,JeLa96b}
strongly suggest that the critical behaviour at $\gamma < 0$ and $\gamma = 0$
belongs to the same universality class. The latter case (Wilson action) is
reflection positive, however. Thus the unitarity violations, if any, seem to
be confined to small (lattice) distance and do not appear on the physical
scale. Of course, this expectation should be tested whenever possible.

\section{Dynamical fermions: Two-state signal}

Here we present our results of dynamical fermion simulations on $6^4$ toroidal
lattice at various values of $\gamma$ and $m_0$, concentrating on $\gamma <
0$. The phase transition in the compact QED with light staggered fermions and
$\gamma < 0$ has been studied last quite some time ago \cite{DaKo88,Ok89}, to
our knowledge.  The two-state signal was observed down to large negative
values of $\gamma=-1.0$.  The availability of the hybrid Monte Carlo fermion
algorithm and larger computer resources suggest the reconsideration of this
transition, as it may be crucial for understanding of QED at strong coupling.
We work on toroidal lattices like \cite{DaKo88,Ok89}.

Expecting that the lattice topology contributes substantially to the two-state
signal, we put emphasis on the comparison of this signal with the analogous
signal in the pure U(1) gauge theory on toroidal lattices.  Therefore we study
$e_P$ in both theories under comparable conditions .  The values of $\Delta
e_P$ are obtained in the following way: At a chosen value of $\gamma$ we made
simulations at several values of $\beta$ in the vicinity of the expected phase
transition. Inspecting the time evolution of $e_P$ the $\beta$ value with
clearest two-state signal was chosen. The values and errors of $\Delta e_P$
have been obtained from a fit of the corresponding histogram with two
Gaussians at the distance $\Delta e_P$.

\begin{figure}[tbp]
  \begin{center}
    \psfig{file=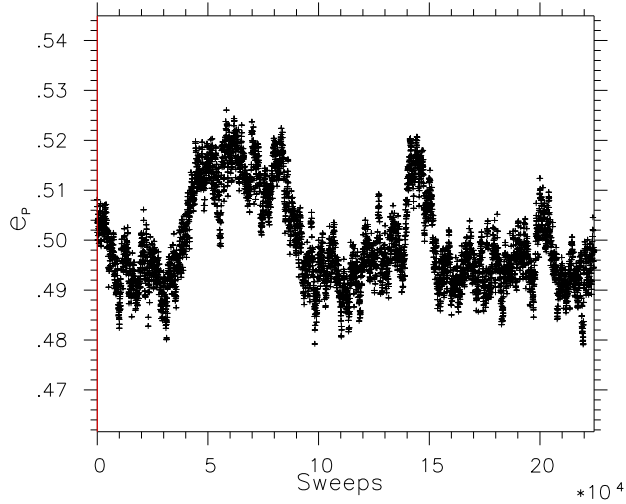,angle=90,width=8.0cm,bbllx=20,bblly=180,bburx=560,bbury=815}
  \end{center}
  \caption{Time evolution of the plaquette $e_P$ at
    $(\beta,\gamma)=(2.08,-1.3)$ and $m_0 = 0.1$ on $6^4$
    lattice. Each point represents an average over 100 measurements.}
  \label{fig:timeevo}
\end{figure}%

The two-state signal in full QED is clearly observable on $6^4$ lattice at
least until $\gamma = -1.5$. The values of $\Delta e_P$ and the corresponding
values of $\beta_c$ on this lattice are collected in table~\ref{tab:gap}.

As an example we show in fig.~\ref{fig:timeevo} the time evolution of $e_P$ in
a run at $(\beta,\gamma) = (2.08,-1.3)$ and $m_0 = 0.1$.  The gap between both
states is $\Delta e_P \simeq 0.018$. The observed tunnelling period indicates
that we need very long observation time: the figure represents $2.2\times10^5$
hybrid Monte Carlo trajectories and the lifetime of the metastable states
is of the order of $4\times10^4$ trajectories. This explains why the two-state
signal was not observed for the same parameters in \cite{Ok89}.  It makes also
clear that increasing the lattice or further substantial decreasing of
$\gamma$ is prohibitively expensive even today.

\begin{figure}[tbp]
  \begin{center}%
    \psfig{file=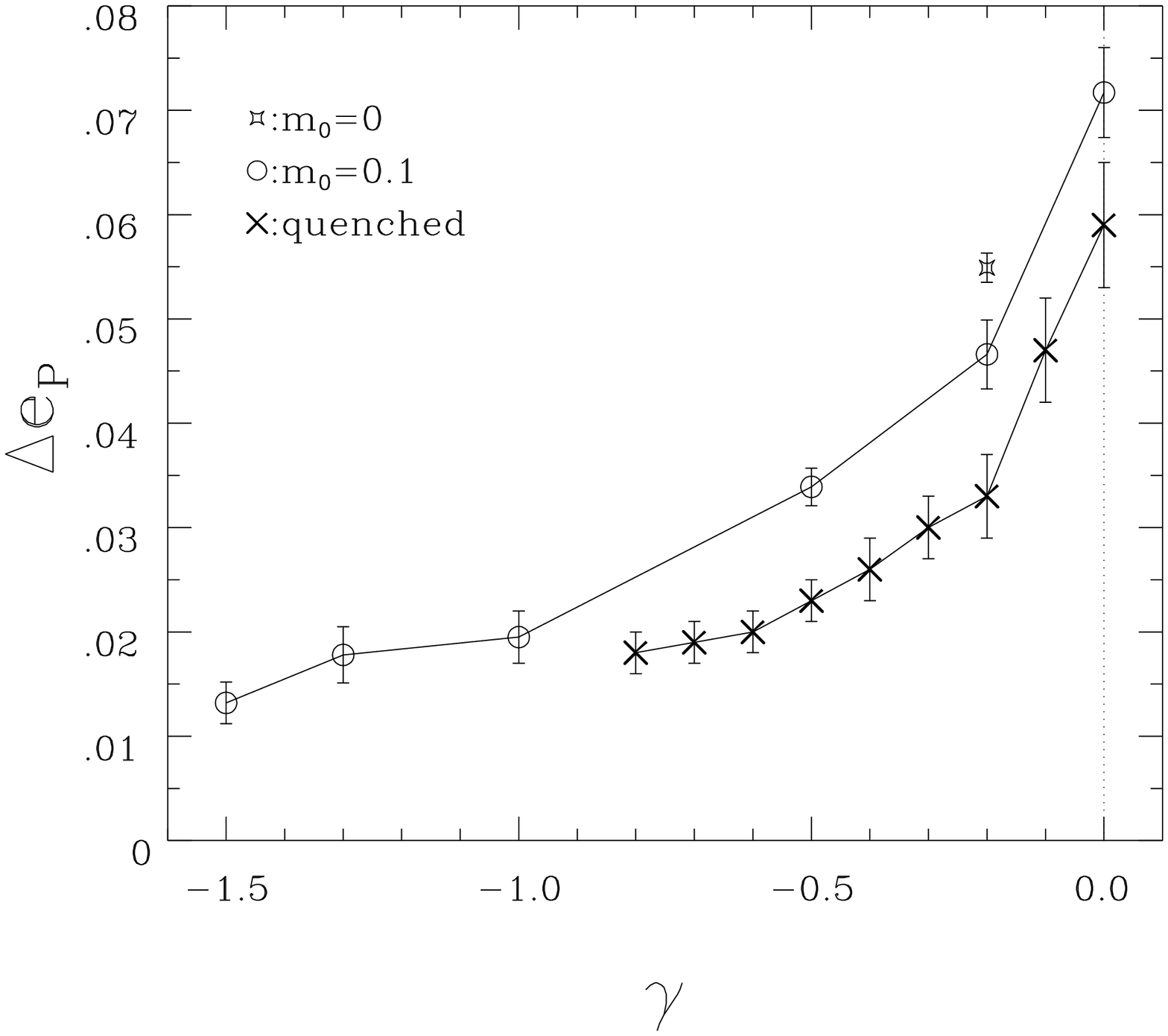,angle=90,width=6.8cm,bbllx=70,bblly=185,bburx=540,bbury=770}
    \psfig{file=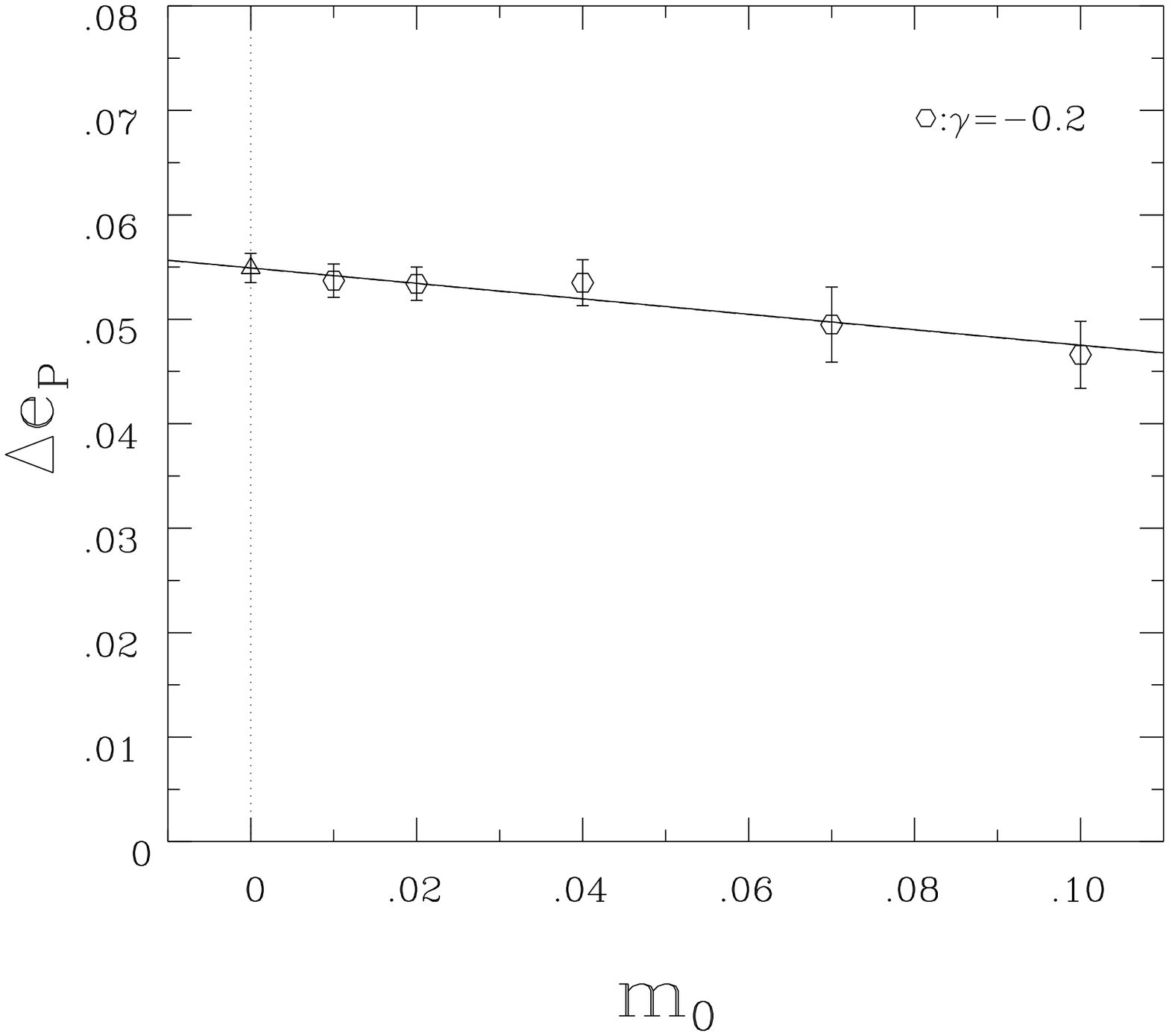,angle=90,width=6.8cm,bbllx=70,bblly=185,bburx=540,bbury=770}%
    \hspace*{-3mm}%
  \end{center}
  \caption{a) Plaquette discontinuity $\Delta e_P$ on $6^4$ lattice as a
    function of $\gamma$ in the full QED for $m_0 = 0.1$ (circles) and in the
    pure U(1) gauge theory (crosses). At $\gamma = -0.2$ also the value of
    $\Delta e_P$ at $m_0 = 0$, obtained by extrapolation, is shown.  b)
    Plaquette gap $\Delta e_P$ on the same lattice as a function of $m_0$ in
    the full QED for $\gamma = -0.2$. A linear extrapolation to $m_0 = 0$ is
    indicated.}
  \label{fig:gap}
\end{figure}%


\begin{table}
  \caption[]{Positions of the transition and the size of the gap $\Delta
  e_P$ on $6^4$ lattice in full QED. }
\vspace{0.2cm}
  \begin{center}
    \begin{tabular}{|c|c|c|c|c|c|c|c|c|}
      \cline{1-4}\cline{6-9}
      \rule[-1.5ex]{0cm}{4.5ex}$\gamma$ &
      $m_0$ & $\beta_{\rm c}$ & $\Delta e_{\rm P}$&&$\gamma$ &
      $m_0$ & $\beta_{\rm c}$ & $\Delta e_{\rm P}$ \\*[1mm]
      \cline{1-4}\cline{6-9}
      0    &0.1 &0.883 &0.072(4) &&-0.2 &0.01 &0.970 &0.054(2)\\
      -0.2 &0.1 &1.015 &0.047(3) &&-0.2 &0.02 &0.977 &0.053(2)\\
      -0.5 &0.1 &1.240 &0.034(2) &&-0.2 &0.04 &0.985 &0.054(2)\\
      -1.0 &0.1 &1.730 &0.020(3) &&-0.2 &0.07 &1.000 &0.050(4)\\
      -1.3 &0.1 &2.080 &0.018(3) &&-0.2 &0.10 &1.015 &0.047(3)\\
      \cline{6-9}
      -1.5 &0.1 &2.350 &0.013(2) &  \multicolumn{5}{c}{} \\
      \cline{1-4}
    \end{tabular}
  \end{center}
\label{tab:gap}
\end{table}

The question is whether the two-state signal is a finite size effect or is
caused by a genuine $1^{st}$ order transition.  Therefore we want to point out
the small size of $\Delta e_P$: At $\gamma = -1.3$ and $m_0 = 0.1$ it is
smaller than that in the pure U(1) gauge theory at $\gamma = -0.5$, where the
$2^{\rm nd}$ order behaviour is clearly seen on spherical lattices
\cite{JeLa96a,JeLa96b}.

Similar observations can be made for $m_0 = 0.1$ in a broad interval
of $\gamma$ between $-0.2$ and $-1.5$. As is seen in
fig.~\ref{fig:gap}a, there the value of $\Delta e_P$ is smaller than
that of the pure U(1) gauge theory at $\gamma = 0$. We have also
observed that $\Delta e_P$ in the pure gauge theory roughly agrees
with that in the model with fermions of small mass, if compared at
identical $\beta$.

The investigation of the $m_0$ dependence of $\Delta e_P$ becomes very
expensive if one approaches small $m_0$, even on the small lattice of size
$6^4$. We have performed such a study at $\gamma = -0.2$, with $m_0$ as small
as $m_0 = 0.01$. The results are shown in fig.~\ref{fig:gap}b. The increase of
$\Delta e_P$ with decreasing $m_0$ is moderate and there is no indication of a
sudden change at very small $m_0$.  A linear extrapolation to $m_0 = 0$,
suggested by the shape of the data, gives $\Delta e_P \simeq 0.055$. A
quadratic extrapolation gives an insignificantly smaller value. This value
still does not exceed the size of the gap in the pure gauge theory at $\gamma
= 0$ on the same lattice.

The comparison of the gap values on $6^4$ lattice in the theory with dynamical
fermions and in pure gauge theory would suggest that even at $m_0=0$ the phase
transition may be already of the $2^{\rm nd}$ order at $\gamma = -0.2$ (since
we have arguments, that the pure gauge theory is really second order already
at $\gamma=0$).  In spite of our expectation that the gap is caused by gauge
fields such a reasoning might be too naive.  Nevertheless, it makes plausible
our hypothesis that the phase transition at $m_0 = 0$ becomes $2^{\rm nd}$
order at some finite $\gamma$, though possibly smaller than $\gamma = -0.2$.

\section{Quenched approximation}
\subsection{Simulations and fermionic observables}
The quenched calculations have been performed at $\gamma = -0.2$ on large
toroidal lattices, typically of the size $16^3 32$, at some points up to $20^3
40$. We used the configurations produced for the measurement of the gauge-ball
masses \cite{CoFr97b}. We avoided the region very close to the phase
transition because of the two-state signal there. This explains a gap in
$\beta$-values in the data and figures below. In this way we also avoided
significant finite size effects. 

The transition point in $\beta$ as determined from the scaling of the
gauge-ball masses in the confinement phase at $\gamma = -0.2$ is
$\beta_c=1.1607(3)$ \cite{CoFr97b}. Our following analysis is crucially
dependent on a good knowledge of $\beta_c$ and will be based on this value. We
have investigated in both phases the chiral condensate, $\sigma$ and $\pi$
susceptibilities and some meson masses. The names of mesons --
fermion-antifermion bound states -- are chosen in analogy with QCD.

Following standard notation we define the chiral condensate
\begin{equation}
  \label{cbc}
  \cbcex = \langle {\rm tr}\, M^{-1} \rangle
\end{equation}
and measure it with a Gaussian noise estimator.
We also determine the
logarithmic derivative $R_\pi$ of the chiral condensate \cite{KoKo93a,GoHo94a}
\begin{equation}
  \label{def_rpi}
  R_\pi = \left.\frac{\partial\ln\cbcex}{\partial\ln
      m_0}\right|_{\beta,\kappa} = \left.
    \frac{m_0}{\cbcex}\frac{\partial\cbcex}{\partial
      m_0}\right|_{\beta,\kappa}\;.
\end{equation}
This may be rewritten as a ratio of zero momentum $\sigma$ and $\pi$ meson
propagators (susceptibilities)
\begin{equation}
  \label{cbc_mes}
  \left.\frac{\partial\cbcex}{\partial m_0}\right|_{\beta,\kappa} =
  C_\sigma(p=0)\;,\quad\quad
  \frac{\cbcex}{m_0} = C_\pi(p=0)\;,
\end{equation}
so that
\begin{equation}
  R_\pi = \frac{C_\sigma(p=0)}{C_\pi(p=0)}\;.
\end{equation}
The first of eqs. (\ref{cbc_mes}) holds in the quenched approximation, if only
the connected part of the $\sigma$ meson propagator is considered. The second
is a Ward Identity \cite{KiSh87}.  The meson operators are defined below.

We expect the validity of the equation of state in analogy with magnetic
systems, \cite{KoKo93a}
\begin{equation}
  \label{eqstate}
  m_0=\cbcex^\delta f(x), \quad x=\frac{t}{\cbcex^{1/\beta_\chi}}, \quad
  t=\beta_c-\beta .
\end{equation}
The suffix $\chi$ of $\beta_\chi$ is added to the usual magnetic exponent
$\beta$ to avoid confusion with the coupling $\beta$.

A non-vanishing chiral condensate requires a positive zero $x_0$ of the
scaling function $f(x)$. The predicted scaling behaviour is
\begin{equation}\label{scalcbc}
  \cbcex= \left\{ \begin{array}{rcl}
    (t/x_0)^{\beta_\chi}
          &\textrm{for}& \beta<\beta_c,\;m_0=0\;,\\
    (m_0/f(0))^{1/\delta}
          &\textrm{for}& \beta=\beta_c,\;m_0\neq 0 \;.
          \end{array}\right.
\end{equation}
The equation of state (\ref{eqstate}) predicts that the ratio $R_\pi$
depends only on the scaling variable $x$,
\begin{equation}
  \label{scalr}
  R_\pi=\left(\delta-\frac{xf^\prime(x)}{\beta_\chi f(x)}\right)^{-1}\;.
\end{equation}

The time\-slice operators for the mesons (projections to zero
momentum) are given by
\begin{equation}
  \label{Omesons}
  {\mathcal O}^{ik} (t) = \sum_{\vec{x}} s^{ik}_{\vec{x} ,t} \chib_{\vec{x} ,t}
  \chi_{\vec{x} ,t} \;,
\end{equation}
with the sign factors~$s^{ik}_x$, $x = \vec{x} ,t$, given in table
\ref{tab:sik}. We measure their correlation functions using point sources and
considering only the connected parts.  Further details can be found e.g. in
\cite{Go86}.

\begin{table}
  \caption[Mesonische Operatoren]%
  {List of measured mesonic operators in the lattice QED with
    staggered fermions. The continuum quantum numbers $J^{PC}$ and the
    name of the corresponding QCD particle are included.  The sign
    factors $s_x^{ik}$ use the phase factors $\eta_{\mu x} =
    (-1)^{x_1+ \cdots +x_{\mu-1}}$, $\zeta_{\mu x} =
    (-1)^{x_{\mu+1}+ \cdots +x_4}$ and $\varepsilon_x = (-1)^{x_1 +
      \cdots + x_4}$. The index $s$ and $a$ at the $J^{PC}$ are for
    the singlet and adjoint representation of the flavor symmetry
    group. The notation follows loosely {\protect\cite{Go86,AlBo93}.}}
  \begin{center}
    \label{tab:sik}
    \begin{tabular}{|c|c|c|c|} \hline
      \rule[-1.5ex]{0cm}{4.5ex}$i$ &
      $s^{ik}_x$ & $J^{PC}$ & particle \\*[1mm] \hline
      1&1&
      $\begin{array}{l} \rule{0cm}{3ex}0^{++}_s \\ 0^{-+}_a \end{array}$ &
      $\begin{array}{l} \sigma \; ({\rm f}_0) \\ \pi^{(1)}
      \end{array}$ \\[5mm]
      \hline
      2&$\eta_{4x}$&
      $\begin{array}{l} 0^{+-}_a \\ 0^{-+}_a \end{array}$ &
      $\begin{array}{l} - \\ \pi^{(2)} \end{array}$ \\[5mm]
      \hline
      3&$\eta_{kx}
      \varepsilon_x \zeta_{kx}$&
      $\begin{array}{l} 1^{++}_a \\ 1^{--}_a \end{array}$ &
      $\begin{array}{l} a_1 \\ \rho^{(1)} \end{array}$ \\[5mm]
      \hline
      4&$\eta_{4x} \eta_{kx} \varepsilon_x \zeta_{kx}$&
      $\begin{array}{l} 1^{+-}_a \\ \rule[-1.5ex]{0cm}{0cm}1^{--}_a
\end{array}$ &
      $\begin{array}{l} b_1 \\ \rho^{(2)} \end{array}$ \\ \hline
    \end{tabular}
  \end{center}
\end{table}

\subsection{Chiral condensate}

\begin{figure}[tbp]
\vspace*{-5mm}
  \begin{center}
    \leavevmode
    \psfig{file=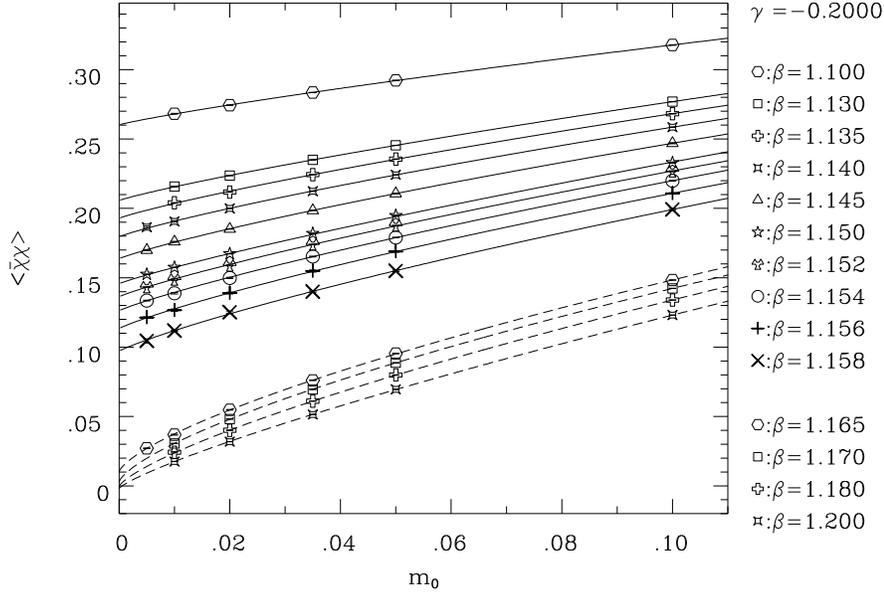,angle=90,width=\hsize}
    \vspace{-5mm}
    \caption{Chiral condensate on the $16^3 32$-lattice as a function
      of $m_0$ for different $\beta$. The lines are our extrapolation
      into the chiral limit according to (\protect\ref{excbc}) (confinement
      phase) and (\protect\ref{excbc_pow}) (Coulomb phase).}
    \label{fig:excbc}
  \end{center}
\end{figure}

\begin{figure}[tbp]
  \vspace*{-5mm}
  \begin{center}
    \leavevmode
    \psfig{file=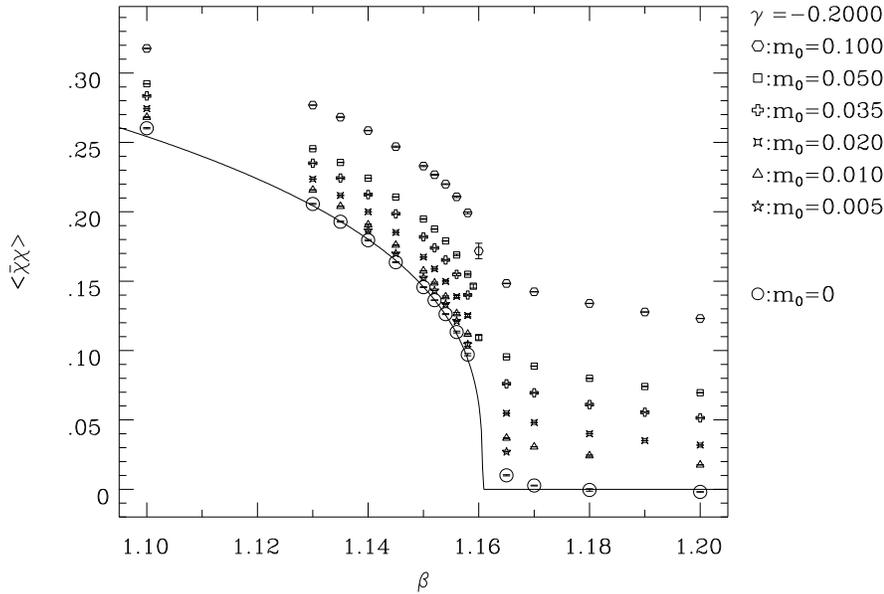,angle=90,width=\hsize}
    \vspace{-5mm}
    \caption{Chiral condensate on $16^3 32$-lattice as a function of
      $\beta$. We show the data for different $m_0$ and our
      extrapolation into the chiral limit (circles). The fit has been done
      according to (\ref{scalcbc}).}
    \label{fig:scalcbc}
  \end{center}
\end{figure}

The data for the chiral condensate are shown in figs.~\ref{fig:excbc} and
\ref{fig:scalcbc}. In both phases the data turned out to be almost
independent of the volume for the $16^3 32$ and larger lattices we used. Figure
\ref{fig:excbc} exhibits the chiral condensate as a function of $m_0$ for
different $\beta$ values. To extrapolate the condensate into the chiral limit
we made for fixed $\beta$ in the broken symmetry phase a fit with the ansatz
\begin{equation}
  \label{excbc}
  \cbcex(m_0) = \cbcex_0 + A m_0 + B m_0 \ln m_0\;,
\end{equation}
motivated by the chiral perturbation theory \cite{GaLe87}. In the Coulomb
phase we use the ansatz
\begin{equation}
  \label{excbc_pow}
  \cbcex(m_0) = \cbcex_0 + C\,(m_0)^z\;.
\end{equation}

The value $\cbcex_0$ is our estimate for the chiral condensate in the
chiral limit. In the confinement phase we get in fact consistent
results by both expressions.  The exponent $z$ is nearly independent
of $\beta$ , being $z\approx 0.8$ in the confinement phase,
significantly different from 1.  In the Coulomb phase the value is
$z\approx 0.7$ close to the transition and rises with $\beta$ to
$z\approx 0.8$ at $\beta = 1.2$.

The results for $\cbcex_0$ are shown in fig.~\ref{fig:scalcbc} as a
function of $\beta$.  A fit $c\,t^{\beta_\chi}$ with $\beta$
approaching the critical coupling $\beta_c$ from below results in the
value of the magnetic exponent $\beta_\chi\simeq 0.33$. The scaling
region has been assumed to lie between $\beta=1.13$ and $\beta=1.16$.
Reducing this interval tends to decrease $\beta_\chi$.

\subsection{Susceptibility ratio $R_\pi$ and equation of state}

The study of $R_\pi$ as suggested in \cite{KoKo93a} allows for a more
sophisticated scaling analysis. As can be seen from (\ref{scalr}), $R_\pi$
should be equal to $1/\delta$ for $\beta=\beta_c$, as long as $m_0$ is small
enough. In the confinement phase with broken chiral symmetry $R_\pi$ vanishes
in the chiral limit, as can be seen from the definition of $R_\pi$. In the
symmetric phase $\sigma$ and $\pi$ meson masses are degenerate, and thus
$R_\pi=1$ in the chiral limit.  The curve of constant $R_\pi$ separates two
groups of curves. It provides an estimate of the critical coupling and
$1/\delta$.

\begin{figure}[tbp]
  \vspace*{-5mm}
  \begin{center}
    \leavevmode
    \psfig{file=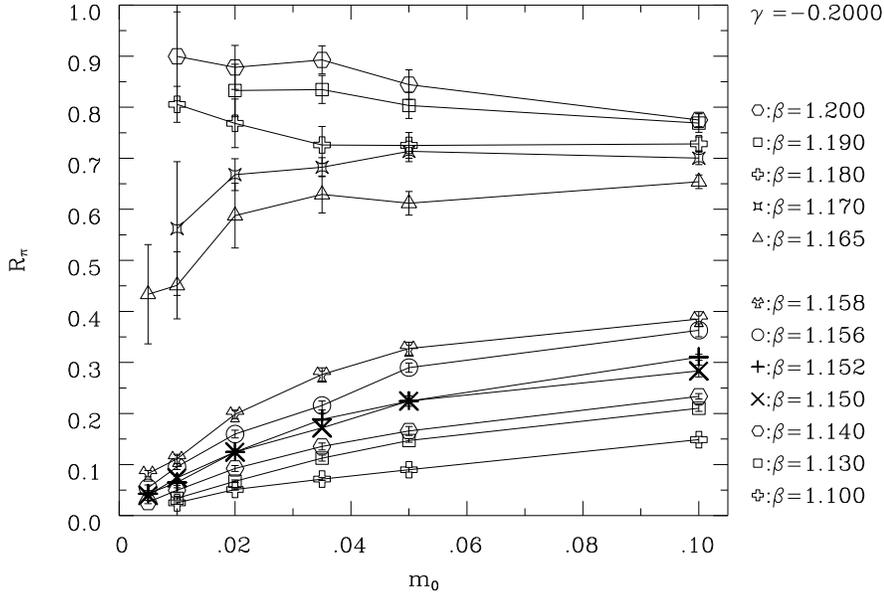,angle=90,width=\hsize}
    \vspace{-5mm}
    \caption{$R_\pi$ as a function of $m_0$ for different $\beta$ on 
      $16^3 32$ lattice. The lines connecting the points at the same
      $\beta$ are drawn to guide the eye.}
    \label{fig:rpi}
  \end{center}
\end{figure}

In both phases the data for $R_\pi$ are again almost independent of
the volume for the $16^3 32$ and larger lattices, thus we show in
fig.~\ref{fig:rpi} the data for $16^3 32$ lattice only. Since no
simulation close to the critical coupling could be done, our estimate
for $\delta$ with this method is bound to have a large uncertainty.
From the value of $R_\pi$ at $m_0 = 0.05$ for the $\beta$-values
closest to the phase transition in each phase we find $\delta =1.7 -
3.3$. The mean-field value would be $\delta=3$.

In the symmetric phase all curves should run towards $R_\pi=1$.  Our data do
not show this behaviour in a clear way.  However, in particular at small
$m_0$, we observe large statistical errors near the transition point. Also,
the expectations of chiral symmetry restoration by mass symmetry of parity
partners may not hold in the quenched approximation. 

\begin{figure}[htb]
  \centerline{
    \hbox{
      \psfig{file=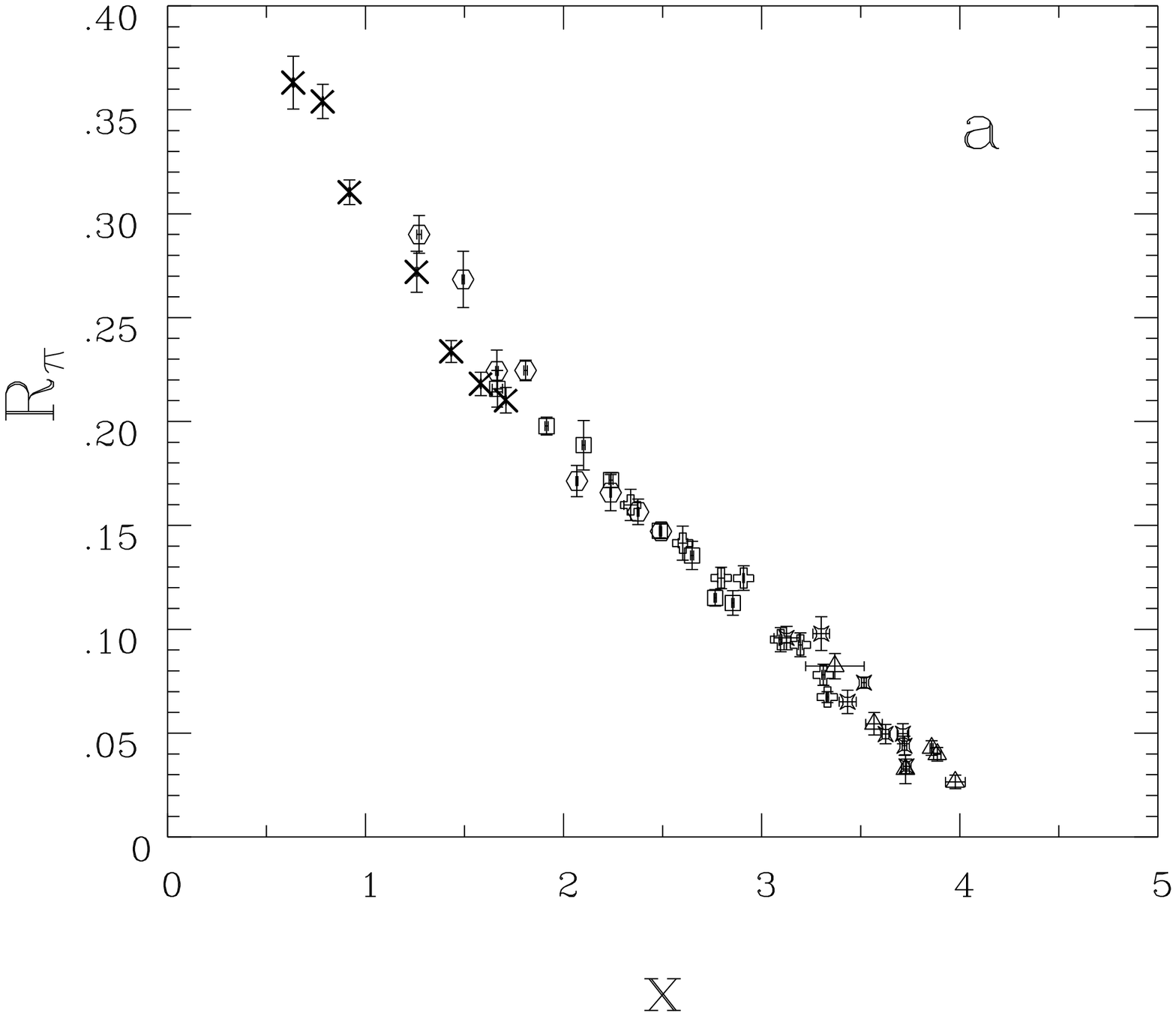,angle=90,width=7cm,bbllx=70,bblly=185,bburx=540,bbury=770}
      \psfig{file=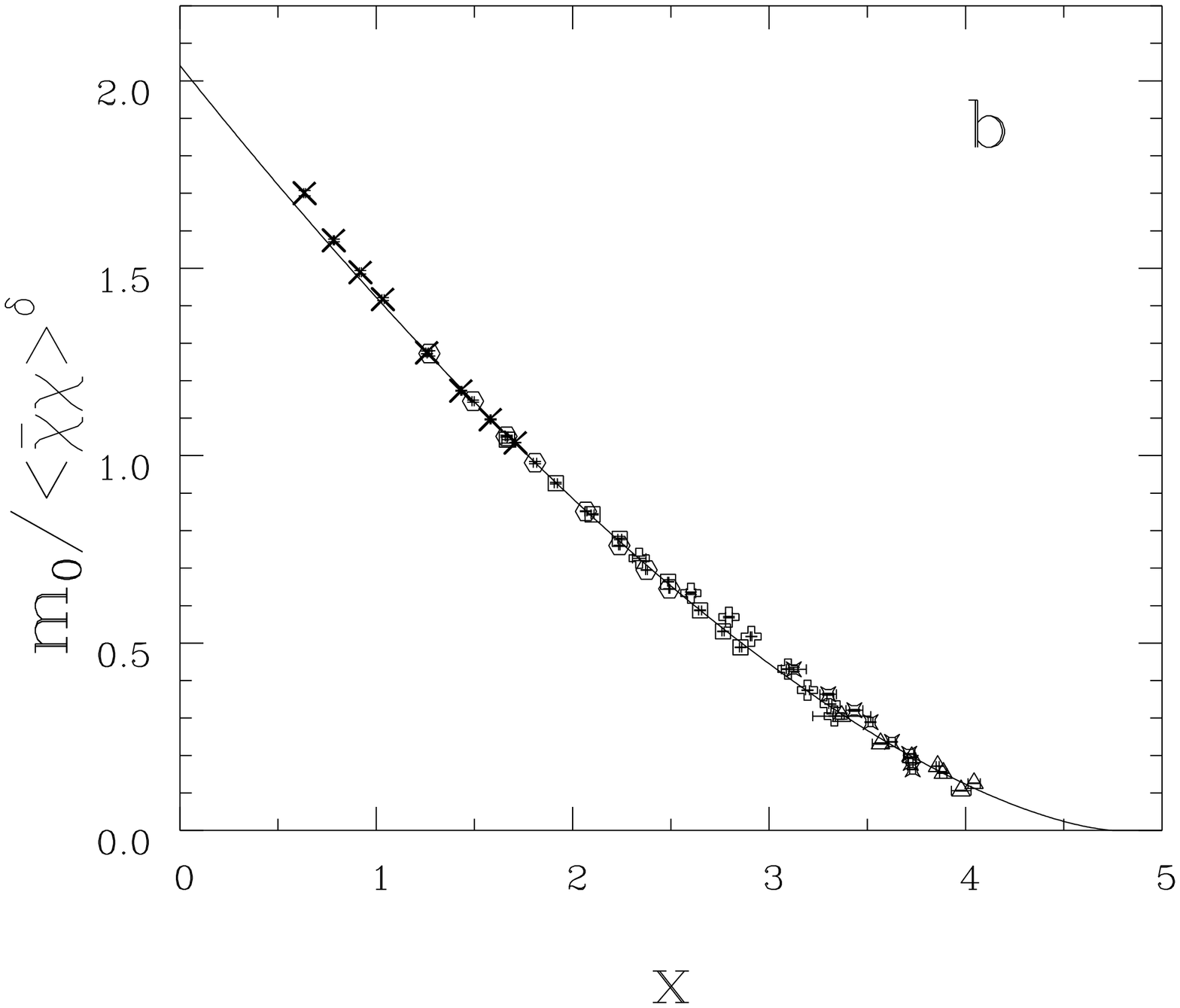,angle=90,width=7cm,bbllx=70,bblly=185,bburx=540,bbury=770}
      }
    }
 \caption{(a) $R_\pi(x)$  and (b) $f(x)=m_0\cbcex^{-\delta}$ as functions
      of the scaling variable $x=t/\cbcex^{-1/\beta_\chi}$ on 
      $16^3 32$ lattice for the exponents $\beta_\chi=0.32$ and
      $\delta=1.82$. The method to obtain these plots and the curve
      in (b) are discussed in the text.}
    \label{fig:findbd}
\end{figure}

To check the validity of the equation of state we also investigate the
dependence of $R_\pi$ on the scaling variable $x$ in the confinement phase,
which is predicted by (\ref{scalr}) to be the only parameter.  This
is done by determining $\beta_\chi$ by the requirement that all data for
$R_\pi$ lie on a single curve depending only on $x$.  We get the best results
for $\beta_\chi=0.32(2)$ (fig.~\ref{fig:findbd}a), which is in good agreement
with the result obtained from the scaling of the chiral condensate. Varying
$\beta_c = 1.1607(3)$ within its error bars neither improves the independence
on $x$ nor changes $\beta_\chi$ significantly.

To determine the exponent $\delta$ more precisely, we use the fact, that
$f(x)$ should be a function of $x$ alone. With (\ref{eqstate}) as an indirect
definition for $f$ and taking our best estimate for $\beta_\chi=0.32$ we tune
$\delta$ so that $m_0\cbcex^{-\delta}$ depends only on $x$.  As shown in
figure \ref{fig:findbd}b, this leads to an excellent description of the data
with $\delta=1.82(10)$. The result for $\delta$ is substantially more precise
than the above attempt to use $R_\pi$ directly.  If this analysis is done with
a slightly different $\beta_\chi$, the obtained results are much worse, and no
unique value of $\delta$ could be obtained.  The scaling function $f(x)$ can
be very well parameterized as $f(x)=0.207\,(4.68-x)^{1.49}$, which is drawn in
fig.~\ref{fig:findbd}b.

Both exponents determined by this method, $\beta_\chi=0.32(2)$ and
$\delta=1.82(10)$, differ from corresponding mean field values. The results
therefore indicate that this model might have a non-Gaussian fixed point.
They may be considered as a first hint that the full compact QED is
nontrivial. However, we cannot give a reliable estimate of the systematic
errors. We note that $\frac{1}{4}\beta_\chi(\delta+1)=0.23(2)$, which does not
agree with the value $\nu=0.35$ as would be expected from the hyperscaling
relation.

\subsection{Meson spectrum}

The masses have been determined with the usual $\cosh$-fits to the
connected propagators \cite{AlGo95}. Each of the four
(cf. table \ref{tab:sik}) correlations functions $C^i(t)$ contains
signals from both parity states (with the only exception of the
pion in the channel $i=2$) and therefore is fitted
to the form
\begin{eqnarray}
B^+ + (-1)^t B^-
&+&\sum_{n=1}^{n^+} A_n^+
\left(e^{-E_n^+ t}+e^{-E_n^+ (L_t-t)}\right)\nonumber\\
&+&(-1)^t\sum_{n=1}^{n^-} A_n^-
\left(e^{-E_n^- t}+e^{-E_n^- (L_t-t)}\right)\;.
\end{eqnarray}
Significant mass determination could be obtained in the $\sigma,
\pi^{(2)}, a_1, \rho^{(1)}$ and $\rho^{(2)}$ channels. For them,
respectively, the numbers of energy levels $(n^+, n^-)$ considered
were in the confinement phase (1,1), (0,2), (1,1), (2,1), (1,2) and in
the Coulomb phase (2,1), (0,2), (2,1), (2,1), (1,2).  The lowest
energy level was interpreted as the mass of the corresponding particle
state. Statistical fluctuations prevent a reliable determination of
$\pi^{(1)}$ and $b_1$.

\subsubsection{Confinement phase}

In the confinement phase finite volume effects become small already at lattice
sizes $12^3 24$; we therefore discuss meson masses obtained on $16^3 32$
lattice only. We note that in this phase the values of both parameters $B^\pm$
obtained by the fits are consistent with zero.

\begin{figure}[tbp]
  \begin{center}
    \psfig{file=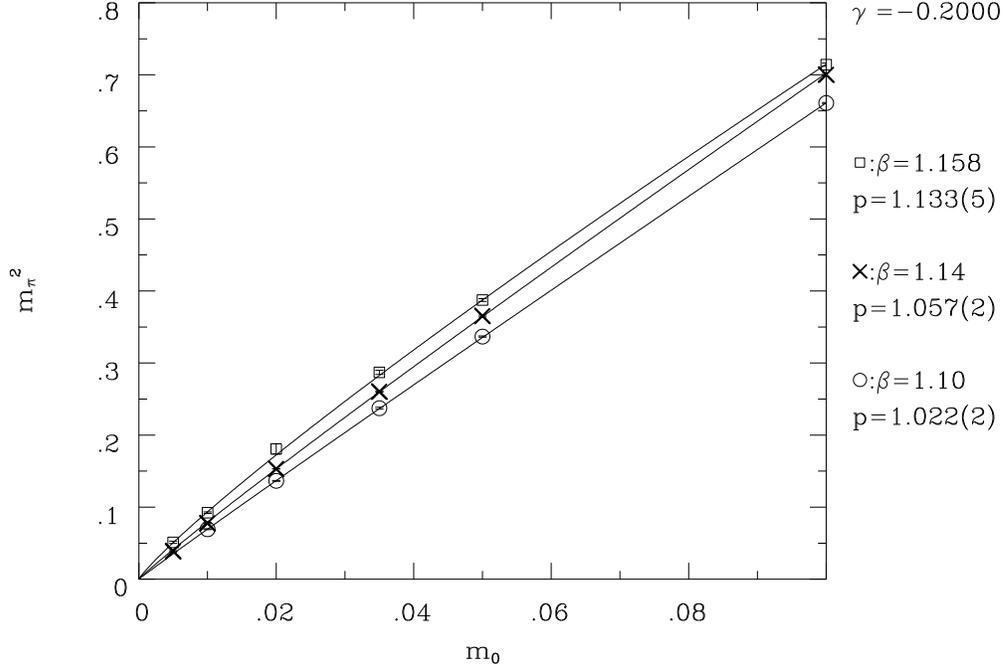,angle=90,width=\hsize,bbllx=70,bblly=50,bburx=540,bbury=770}
  \end{center}
  \caption{The figure shows $m_\pi^2$ vs. $m_0$ for $\beta=$
    1.10 (circles), 1.14 (crosses) and 1.158 (squares); the fit $(m_\pi)^2
    \propto (m_0)^p$ gives values of $p$ close to one.}
  \label{fig:pi.m0}
\end{figure}

Close to the chiral limit one expects
\begin{equation} \label{PCAC}
(m_\pi)^2 \propto m_0\;.
\end{equation}
In quenched studies there may be corrections due to so-called chiral
logarithms \cite{BeGo92,Sh92,Ma96}, thus we fit our results (fig.
\ref{fig:pi.m0}) to $(m_\pi)^2 \propto (m_0)^p$. We find values of $p$ close
to one (between 1.02 and 1.13 close to the phase transition).

\begin{figure}[tbp]
  \begin{center}
    \psfig{file=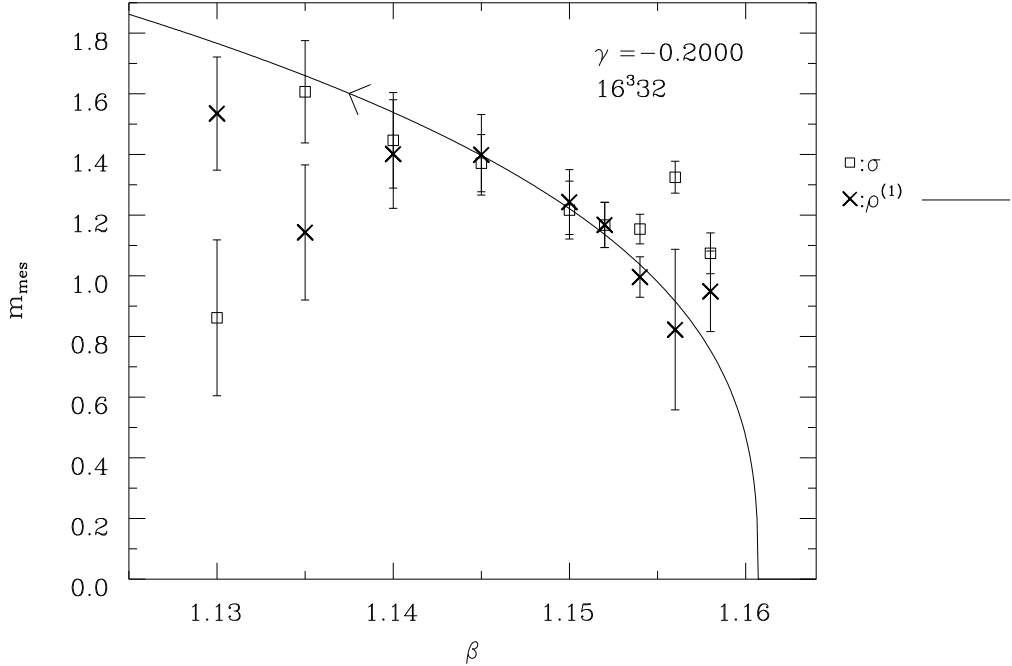,angle=90,width=\hsize,bbllx=70,bblly=50,bburx=540,bbury=770}
  \end{center}
  \caption{Masses of $\sigma$ (squares) and $\rho^{(1)}$ (crosses)
    extrapolated to $m_0 = 0$. The curve is $m = c\, t^{0.35}$, with fitted
    $c$.}
  \label{fig:mesall}
\end{figure}

The $m_0$ dependence of the non-Goldstone state masses has been parameterized
linearly, $m \propto m_0 + const$, as usual.  Fig.  \ref{fig:mesall} gives the
values of $\sigma$ and $\rho^{(1)}$ as they emerge from the extrapolation to
$m_0=0$. The mass of $\rho^{(2)}$ is compatible with that of $\rho^{(1)}$. The
$a_1$ is heavy ($m>1.5$) and seems not to scale at the phase transition.

The data is not good enough to warrant a fit to a critical exponent.
However, assuming the non-Gaussian value $\nu=0.35$ obtained in
\cite{JeLa96a,JeLa96b,CoFr97b} and the knowledge of $\beta_c$, a fit
$m = c\, t^\nu$ of the $\rho^{(1)}$-mass is shown in the plot fig.
\ref{fig:mesall} to be compatible with the data.  The $\sigma$ has
mass values of the same order of magnitude as the $\rho$ whereas the
$a_1$ is definitely heavier by almost a factor of 2.

Comparing the $\rho$ and $\sigma$ with the the gauge-balls (determined from
the same runs in \cite{CoFr97b}) we find meson masses quite close to the
$T_1^{+-}$ gauge-ball group. For example, the amplitude of $\rho$ is $c=
6.0(2)$ and the corresponding $T_1^{+-}$-gauge-ball amplitude is 5.4(3).

\subsubsection{Coulomb phase}

In the Coulomb phase we expect and we do find sizeable finite volume
dependence, presumably caused by massless photons and light fermions.  The
mesons are therefore expected to be resonances. This is like in the recent
study of gauge-ball masses in the Coulomb phase \cite{CoFr97b}. The volume
dependence there was explained as a signal of multi-photon states, but a
reliable quantitative extrapolation to the thermodynamic limit was not
possible.  For mesons the situation is even worse. In particular the Bohr
radius, proportional to the inverse fermion mass (which we have not
determined), might be much larger than the lattice. Therefore our
investigation of the spectrum in the Coulomb phase has only an exploratory
character, checking whether unexpected phenomena might be seen in the Coulomb
phase close to the phase transition to the confinement phase.

\begin{figure}[tbp]
  \begin{center}
    \psfig{file=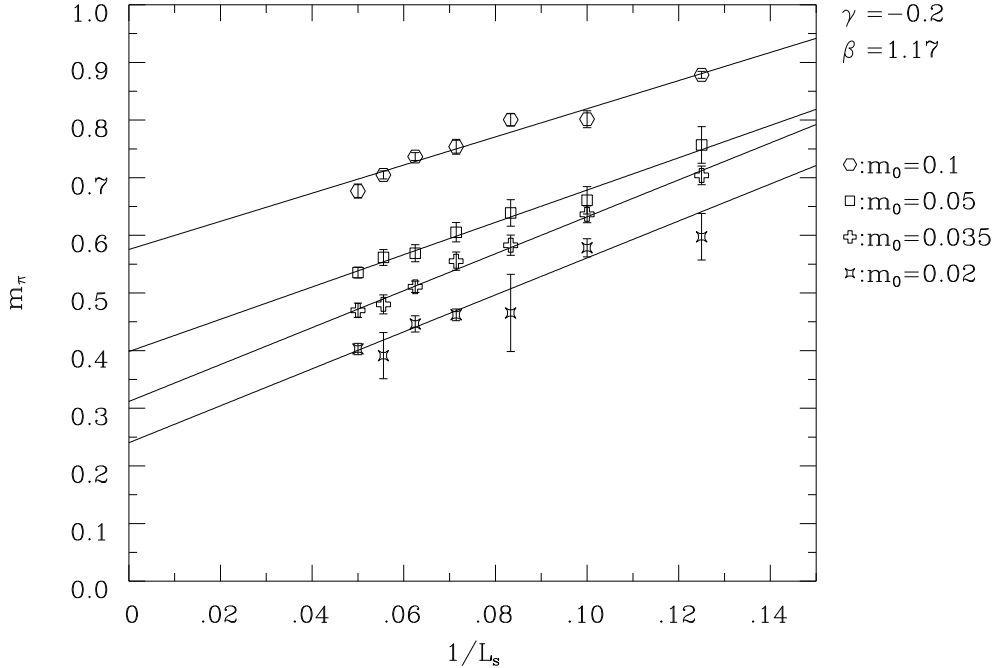,angle=90,width=\hsize,bbllx=70,bblly=50,bburx=540,bbury=770}
  \end{center}
  \caption{The $\pi$ mass in  the Coulomb phase at $\beta=1.17$
plotted vs. $1/L_s$ for various $m_0$.}
  \label{fig:mes1}
\end{figure}

\begin{figure}[tbp]
  \begin{center}
    \psfig{file=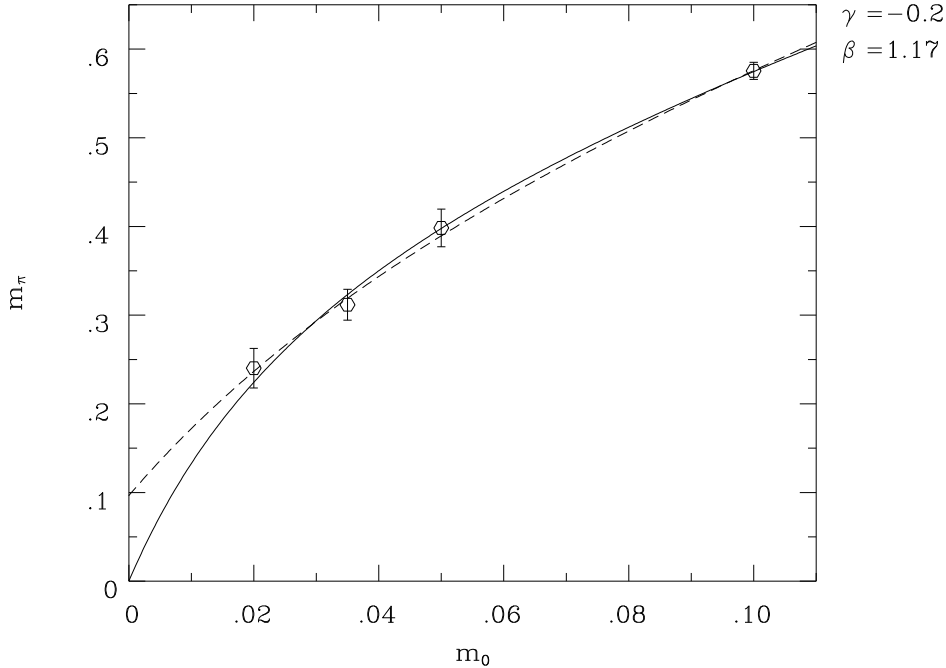,angle=90,width=\hsize,bbllx=70,bblly=50,bburx=540,bbury=770}
  \end{center}
  \caption{The values of $m_\pi$ extrapolated to infinite volume against
    $m_0$. We also show the results of fits (\ref{m_pi}) with (solid line)
and without constraint $A=0$. }
  \label{fig:mes2}
\end{figure}

Fig.~\ref{fig:mes1} indicates, that the size dependence may follow a
linear $1/L_s$ behaviour. Using the correspondingly extrapolated
values for the pion (para-onium) mass, we obtain fig.~\ref{fig:mes2}.
For increasing fermion masses one expects $m_\pi \simeq 2 m_0$, i.e.
twice the mass of the constituents. For small constituent mass, one
would expect $m_\pi \propto m_0$ like the positronium spectrum.
However, it is not completely excluded that the massless QED in the
Coulomb phase at strong coupling shows some nonperturbative phenomena
due to IR singularities.

Therefore we have fitted the mass values with a simple three-parameter
curve
\begin{equation}
 m_\pi=\frac{A+ B m_0+2 m_0^2}{C+m_0}
\label{m_pi}
\end{equation}
either keeping $A$ fixed to 0 (solid line in fig.~\ref{fig:mes2}) or letting
$A$ free (dashed line, $A=0.1(1)$).  We found, that the data are compatible
with both, vanishing and non-vanishing values of $m_\pi$ at $m_0 = 0$, and the
conventional expectation of vanishing meson masses is in agreement with the
data.  The non-vanishing value would imply a decoupling of the corresponding
state in the continuum theory if we require that a continuum limit should be
taken at $\beta \to \beta_c$ and $m_0 \to 0$.

Comparing the mass values for all states as obtained on a fixed lattice size
$16^3 32$ we observe an approximate mass degeneracy of the chiral partners
$\sigma$ and $\pi$ even at finite $m_0$, consistent with chiral symmetry in
the $m_0 = 0$ limit.  This is in agreement with the discussion of $R_\pi$
approaching 1.  Also $\rho^{(2)}$ and $a_1$ masses are close to each other and
to the $\sigma$ and $\pi$ masses. Thus the data do not show any significant
fine or hyperfine splitting.

This explorative study of the spectrum indicates that much better data and
understanding of finite size effects are required to obtain results for the
spectrum in the Coulomb phase which would bring further insight.

%

\section{Summary}

We have studied the compact U(1) gauge theory with staggered fermions
on toroidal lattices on both sides of the phase transition line in the
$(\beta, \gamma)$-plane.
\begin{itemize}
\item With dynamical fermions the observed gap $\Delta e_P$ on
  $6^4$-lattices increases with respect to the pure gauge theory,
  decreases with decreasing $\gamma$ but does not vanish in the
  considered domain $-1.5\leq \gamma\leq 0$.  At $\gamma=-0.2$ and for
  massless fermions it is of the same size as for the pure gauge
  theory at $\gamma=0$.  Since there the gap disappears on sphere-like
  lattices we conjecture, that also the fermionic theory may in fact
  have a critical phase transition at negative $\gamma$.
\item On large lattices we study the chiral condensate and the meson spectrum
  in the quenched approximation. Quenched results should not be
  over-interpret\-ed.  We hope, however, that these results stimulate further
  studies of the compact lattice QED with dynamical fermions.
\item The chiral condensate is determined with high precision.  The main
  result of our study is the observation that the chiral condensate can be
  very well described as a function of both parameters, $\beta$ and $m_0$, by
  means of an equation of state with the relevant critical indices that have
  non-Gaussian values: $\beta \simeq 0.32$ and $\delta \simeq 1.8$.
\item In the confinement phase we observe chiral symmetry breaking and
  Goldsto\-ne-boson behaviour of the pion, as expected. The mass squared obeys
  a PCAC behaviour. All masses (extrapolated to vanishing bare fermion mass)
  have a behaviour compatible with a non-Gaussian scaling at the phase
  transition in $\beta$.
\item The results for masses in the confinement phase are generally of better
  statistical quality than the Coulomb phase results. In the Coulomb phase the
  masses of the pion and other states are of comparable size.
\end{itemize}

We obtained a better insight into some aspects of compact lattice QED,
in particular concerning chiral symmetry breaking and the meson
spectrum in the confinement phase. However, higher statistics will be
necessary to improve the quality of some of our results in this phase
and to determine the scaling behaviour of the meson masses. In the
Coulomb phase not only considerably higher statistics, but also
understanding of finite size effects of various origin is required.
Here the problems are similar to those of the noncompact QED, except
that $\beta_c$ can be determined from the other, technically simpler
phase. This might be an advantage of the compact formulation.

\begin{ack}
  We thank M.~G\"ockeler for reading the manuscript and useful
  comments and M.-P. Lombardo for discussions. The computations have
  been performed on the CRAY-YMP and CRAY-T90 of HLRZ J\"ulich. J.C.,
  W.F. and J.J. thank HLRZ, and T.N. thanks SCRI for the hospitality.
  The work was supported by DFG.
\end{ack}


\bibliographystyle{wunsnot}   


\end{document}